\documentclass[final,12pt,3p]{elsarticle}
\usepackage{latexsym}
\usepackage{amssymb}
\usepackage{amsmath,bm}
\usepackage{cancel}
\usepackage{color}
\definecolor{red}{rgb}{1.0,0.0,0.0}
\definecolor{blue}{rgb}{0.0,0.0,1}
\definecolor{green}{rgb}{0.29, 0.33, 0.13}
\newcommand{\ket}[1]{\left| #1 \right>} 
\newcommand{\bra}[1]{\left< #1 \right|} 
\newcommand{\meanv}[1]{\left< #1 \vphantom{#1} \right>} 
 
\begin{document}
\begin{frontmatter}
\title{
Formation of spectral triplets induced by parity deformation in a quantum dot-cavity system}
\author[label1]{Luisa M. Hurtado-Molina}
\author[label2]{Saravana Prakash Thirumuruganandham}
\author[label1]{Santiago Echeverri-Arteaga}
\author[label1]{Edgar A. G\'omez\corref{cor1}}
\address[label1]{Programa de F\'isica, Universidad del Quind\'io, 630004, Armenia, Colombia\fnref{label4}}
\address[label2]{Department of Technological Operations and High Performance Computing, Empresa P\'ublica SIEMBRA EP
Centro de Emprendimiento, Urcuqu\'i, Imbabura 100115, Ecuador}
\cortext[cor1]{Corresponding author}
\ead{eagomez@uniquindio.edu.co}
%
\begin{abstract}
In this work, we conduct an investigation on the optical emission properties of a pumped-dissipative QD-cavity system within the Lindblad master equation approach, where the cavity mode is modeled through the R-deformed realization of the Heisenberg algebra (RDHA). We found that the parity deformation parameter associated with the RDHA gives rise to a collective phenomenon by reordering the one-photon optical transitions of the system, and it is demonstrated that there are two critical values for this parameter where a new type of spectral triplet arises in the emission spectrum. Moreover, our results reveal that depending on the critical value of the deformation parameter, the spectral triplets observed in the emission spectrum exhibit signatures of the weak- and strong-coupling regimes simultaneously.   
\end{abstract}
\end{frontmatter}
%
\section{Introduction}\label{intro}
\noindent
Within the framework of cavity quantum electrodynamics (cQED), the Jaynes-Cummings (JC) model remains to be the fundamental model for the quantum description of light-matter interaction. In spite of its simplicity, this theoretical model has been used extensively and quite successfully in the description of several quantum phenomena. For example, in the understanding of the collapse and revival of Rabi oscillations in quantum-dot cavity systems (QD-cavity)~\cite{Dory:2016,Michler:2009,Cummings:1965,Eberly:1980}, the phenomenon of light-matter entanglement and dynamics of atomic population inversion~\cite{Bose:2001,Scheel:2003}. Furthermore, the JC model has been used in studies on generation of Schr\"odinger cat states of the quantized field~\cite{Yurke:1986, Moya-Cessa:1995} as well as in the production of nonclassical states of light and atoms~\cite{Ghosh:1997,LoFranco:2007,Dodonov:2002}. It has also been demonstrated to be useful for designing and realization of possible devices with applications in quantum information processing and quantum computation~\cite{Azuma:2011,Moreau:2011,Tsintzos:2008,Wei:2014}. Recent investigations on quantum networking by light-matter interfacing are suggested the JC model as a promising for the development of distributed quantum networks on larger scales~\cite{Reiserer:2015,Pomorski:2019} and new architectures envisaging the future quantum internet~\cite{Durl:2017}. Surprisingly, the JC model serves as an elementary "building block" towards the understanding of novel phenomena in more complex quantum systems, e.g. many-body phenomena in QED-cavity arrays~\cite{Tomadin:2010}, multipartite entanglement in  semiconductor quantum-dots~\cite{Xue:2012}, phase transition of light~\cite{Koch:2009} and
Mott-insulator-to-superfluid transition in strongly correlated polaritons~\cite{Schmidt:2010,Hartmann:2006}. Due to the success of the JC model in the description of important quantum phenomena, it has inspired interesting theoretical works in various directions. For example, the use of theoretical approaches such as the algebraic operator method~\cite{Sukumar:1984,Karasev:1991}, super-group~\cite{Kochetov:1992,Gieres:1997} and super-algebras approaches~ \cite{Buzano:1989} for finding exact solutions to cQED systems. On the other hand, some extensions to the boson oscillator algebra have been given through the deformation of the usual commutation relation of the field operator, which are known as the $q$-deformed~\cite{Chaichian:1990, Manko:1993}
 and $f$-deformed oscillator formalism~\cite{Manko:1997, Sanchez:2012} and they have also been incorporated in the JC model. In particular, these $q$- or $f$-deformed oscillator approaches have been useful in the description of supersymmetric and shape-invariant systems~\cite{Aleixo:2012, Aleixo:2000}. Another case of interest within the deformed algebra approaches and not related to the $q$- or $f$-deformed formalism is known as the R-deformed Heisenberg algebra (RDHA) which has found utility in the characterization of nonclassical properties of light~\cite{Mojaveri2015, Mojaveri:2017, Dehghani:2019}. Interestingly, this R-deformed algebra has been useful in the description of para-fields~\cite{Greenberg:1965}, para-statistics~\cite{Govorkov:1983} and the solution of essential quantum systems such as the pseudo-harmonic~\cite{Dong:2002} oscillator and the Calogero model~\cite{Meljanac:2002}. Also, it has been demonstrated that the RDHA is a powerful tool to deal with fractional spin fields in supersymmetric quantum mechanics~\cite{Plyushchay:1996} and generalized oscillator systems within the quantum field theory~\cite{Ohnuki:1982}. The RDHA is attracting considerable interest to the quantum optics community and mainly when it is considered in the JC model through the replacement of the boson operator by the R-deformed partner. For example, theoretical investigations related to the RDHA and the degree of entanglement~\cite{Dehghani:2016}, as well as in the atomic emission spectrum and variation of the geometric phase~\cite{Altowyana:2020} in the deformed--JC model. The aim of this work is twofold: on the one hand, to investigate the effect that has the parity operator belonging to the RDHA on the optical transitions of a pumped-dissipative QD-cavity system. On the other hand, to characterize the emitted light by the quantum system that is feasible experimentally. This paper is organized as follows. In Section~\ref{sec:TModel}, we present the model that describes the pumped-dissipative QD-cavity system within the framework of RDHA. In Section~\ref{sec:results}, we present a detailed analysis of our numerical results. Finally, the concluding remarks appear in Section~\ref{sec:conclusions}.
\section{Theoretical model}~\label{sec:TModel}
\noindent
The most widely used theoretical model for describing the interaction between light and matter is known as the Jaynes-Cummings (JC) model. It describes a quantum emitter (QE) interacting with an electromagnetic cavity mode that in the rotating wave approximation its Hamiltonian is given by ($\hbar=1$)
\begin{equation}\label{JCmodel}
 \hat{H}_\lambda=\frac{{\omega}_c}{2}\{\bm{\hat{a}},\bm{\hat{a}}^{\dagger}\}+\frac{{\omega}_x}{2}\hat{\sigma}_z+g(\bm{\hat{a}}^{\dagger}\hat{\sigma}
 +\bm{\hat{a}}\hat{\sigma}^{\dagger}),
\end{equation}
where $g$ defines the light-matter interaction constant between the cavity mode and the QE. Additionally, $\omega_{x}$ and $\omega_{c}$ are the frequencies associated with the QE and the cavity mode. Moreover, $\delta=\omega_x-\omega_c$
defines the detuning of the QE from the cavity resonance. Notice that here $\bf{\hat{a}^{\dagger}}$ and $\bf{\hat{a}}$ are creation and annihilation operators within the framework of RDHA. Moreover, it is worth mentioning that these R-deformed operators satisfy the (anti-) commutation relations:
$\{\hat{R},\bm{\hat{a}}\}=0$, $\{\hat{R},\bm{\hat{a}}^{\dagger}\}=0$, $[\bm{\hat{a}},\bm{\hat{a}}^{\dagger}]=\hat{I}+2\lambda\hat{R}$ as well as the properties $\bm{\hat{a}}^{\dagger}\bm{\hat{a}}=\hat{N}+\lambda(\hat{I}-\hat{R})$ and $[\hat{R},\hat{N}]=0$~\cite{Mojaveri2015}. The number operator $\hat{N}$ has been introduced in the RDHA to be different from the traditional product in quantum mechanics, but follows similar rules such as $[\hat{N},\bm{\hat{a}}]=-\bm{\hat{a}}$ and $[\hat{N},\bm{\hat{a}}^{\dagger}]=\bm{\hat{a}}^{\dagger}$.  
Additionally, $\lambda$ is the so-called Wigner’s deformation parameter or parity deformation parameter and $\hat{R}$ defines the parity operator that satisfies the following properties
$\hat{R}^{2}=\hat{I}$,  $\hat{R}^{\dagger}=\hat{R}^{-1}=\hat{R}$, moreover the action of the set of operators $\{\bm{\hat{a}}^{\dagger}, \bm{\hat{a}},\hat{N},\hat{R}\}$ on normalized eigenfunctions are given by
\begin{align}
\hat{N}\ket{n}&=n\ket{n},\\
\hat{R}\ket{n}&=(-1)^{n}\,\ket{n},\\
\bm{\hat{a}}\ket{n}&=\sqrt{n+2\lambda\xi_n}\,\ket{n-1},\\
\bm{\hat{a}}^{\dagger}\ket{n}&=\sqrt{n+1+2\lambda\xi_{n+1}}\,\ket{n+1}\,\,\,\,\,\, \text{where} \qquad 
\xi_n \,=\, 
\begin{cases}
            1 &         \text{if }\: n\,\text{ is odd}\\
            0 &         \text{if }\: n\,\text{ is even.} \end{cases}
\end{align}
On the other hand, we denote by $\hat{\sigma}=\ket{G}\bra{X}$ the lowering operator for the QE, such that the action of the lowering operator on the excited state $\ket{X}$ leads to the ground state $\ket{G}$. It is worthwhile noting that the deformed--JC model has a conserved quantity that is associated with the total number of excitations of the system and it is defined through the operator $\hat{\mathcal{N}}_{exc}=\hat{N}+\hat{\sigma}_z/2
$ that is diagonal in the bare-states basis $\big\{\ket{\alpha,n}\equiv\ket{\alpha}\vert_{\alpha=G}^{X}\otimes\ket{n}\vert_{n=0}^{\infty}\big\}$. In this basis of states, $n$ and $\alpha$ corresponds to the number of photons in the cavity and one of the two possibles states of the QE, respectively. Since the operator $\hat{\mathcal{N}}_{exc}$ defines the conserved quantity ($[\hat{\mathcal{N}}_{exc}, \hat{H}_{\lambda}]=0$) it is possible to study the quantum dynamics of the coupled system within separate subspaces of the full state-space. In fact, each eigensubspace is labeled by the corresponding eigenvalue
$n=0,1,2,...$ of $\hat{\mathcal{N}}_{exc}$ and it is frequently called the $n$-th rung in the JC ladder of states. In other words, the Hilbert space
of the coupled system $\hat{H}_\lambda$ is decomposed into a direct sum of invariant subspaces as $\hat{H}_\lambda=\hat{H}_0\bigoplus_{n=1}^{\infty}\hat{H}_n$, where the subspace $\hat{H}_0$ is spanned only by the state $\ket{G,0}$ that corresponds to the null-excitation subspace and the single-excitation
subspace $\hat{H}_n$ is spanned by states $\{\ket{G,n},\ket{X,n-1}\}$. As a result of this decomposition is that the matrix representation of
$\hat{H}_\lambda$ becomes block diagonal with blocks given by
\begin{eqnarray}
\hat{H}_{0}&=&\frac{\delta}{2}+\lambda,\\
\hat{H}_n&=&\begin{pmatrix}\label{eqnth}
(n+\lambda)\omega_c-\frac{\delta}{2} & g \sqrt{n+2\lambda \xi_n} \\[4pt]
g\sqrt{n+2\lambda\xi_n} &(n+\lambda)\omega_c+\frac{\delta}{2}
\end{pmatrix}.
\end{eqnarray}
where each $H_n$ is two-dimensional.
\section{Results and discussions}\label{sec:results}
\begin{figure}[ht!]
\centering
\includegraphics{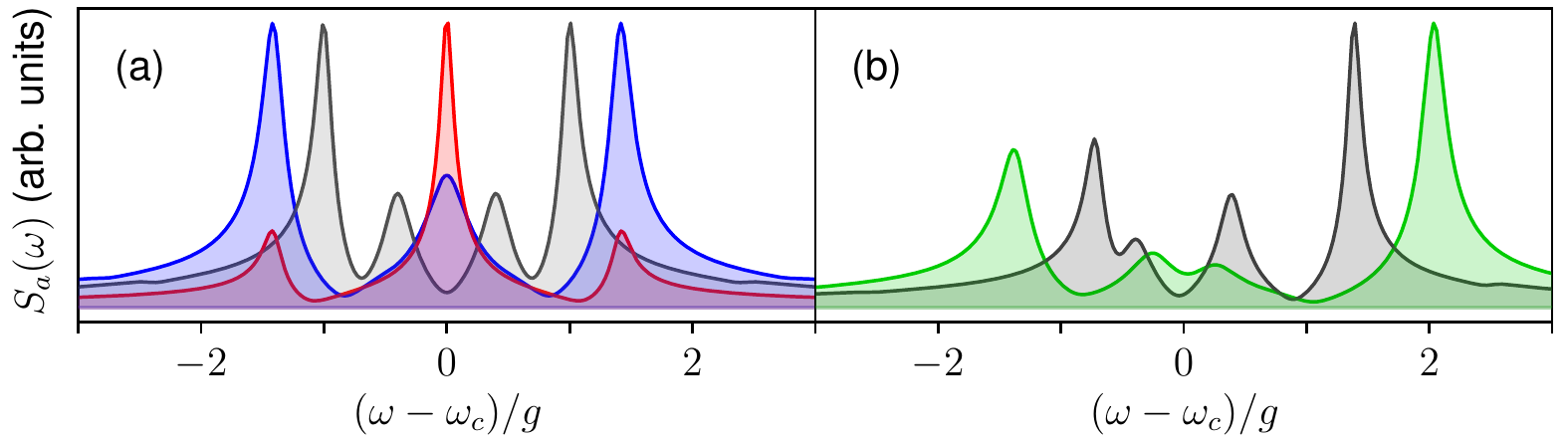}
\caption{Panel (a) shows a comparison of the PL spectrum of the system at resonance ($\delta=0$) for different values of the deformation parameter $\lambda$. The non-deformed case ($\lambda=0$) is shown as solid-gray line, whereas the at the critical values of deformation parameter $\lambda=0.5$ as solid-blue line and $\lambda=-0.5$ as solid-red line. Panel (b) shows a comparison of the PL spectrum when the system is detuned ($\delta/g=0.7$) from the resonance. The non-deformed case is shown as solid-gray line and for the deformed case $\lambda=0.9$ is shown as solid-green line. The other parameters are fixed to $\kappa/g=0.083$, $\gamma/g=0.017$ and $P/g=0.05$. }\label{fig:PL}
\end{figure}

To investigate the relation between the optical transitions of the coupled system and the deformation parameter $\lambda$, we obtain the eigenvalues and eigenvectors associated to the matrix given by Eq.~(\ref{eqnth}) that corresponds to the $n$-th rung in the JC ladder. This methodology allows to obtain the energies for a particular number $n$ of excitations as follows: 
\begin{equation}\label{Eigenv} 
E_{n\pm}= \omega_c(n+\lambda)\pm\frac{\mathcal{R}_n^\lambda}{2}
\end{equation}
where $
\mathcal{R}_n^\lambda=\sqrt{4g^{2}(n+2\lambda\xi_n)+\delta^{2}}$ defines the generalized Rabi frequency, and two new eigenstates $\ket{n\pm}$ which are known as dressed states. It is worth to mention that for the case of the non-deformed JC model ($\lambda=0$) and without dissipation the dressed states are energetically split by $2g\sqrt{n}$ at resonance ($\delta=0$) which is a manifestation of the quantization in the system. In particular, the optical transitions obtained from the difference in energy of two consecutive rungs of the JC ladder accounts for the peaks observed in the photoluminescence (PL) spectrum. The allowed one-photon optical transitions could be classified as those that produce an infinite sequence of peaks --inner Rabi doublets-- positioned at distances of $\pm g(\sqrt{n}-\sqrt{n-1})$ in the PL spectrum and they are piling up towards zero from the first Rabi doublet ($n=1$). In contrast with these optical transitions, there are also another ones that produce peaks --outer Rabi doublets-- positioned at distances of $\pm g(\sqrt{n}+\sqrt{n-1})$ that goes beyond the first Rabi doublet whose predominance in the PL spectrum is barely noticeable and therefore they are not considered in the present analysis. 
\begin{figure}[ht!]
\centering
\includegraphics[scale=1]{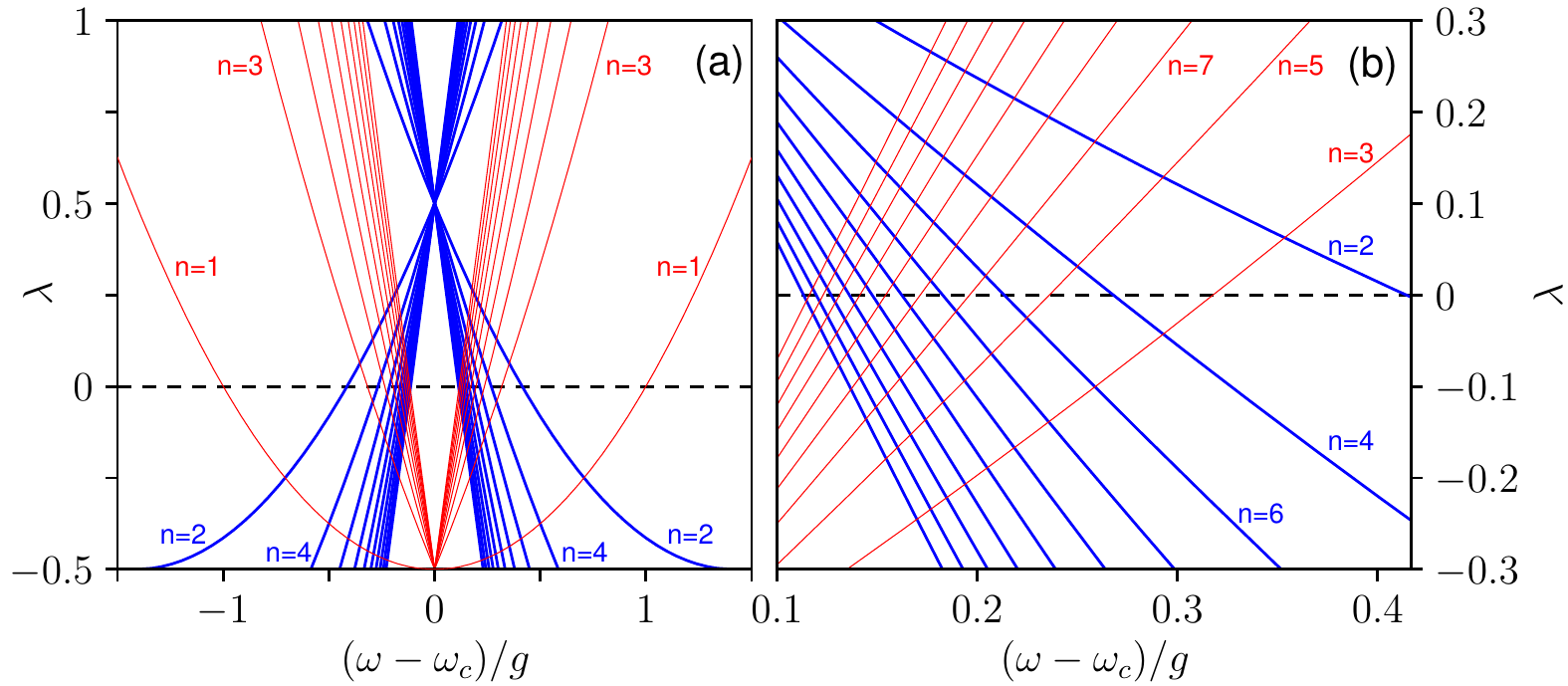}
\caption{Panel (a) shows the spectral position for the first  inner Rabi doublets ($n=1,2,3...19$) as a function of the parity deformation parameter $\lambda$. The solid-blue lines correspond to the even--transitions and the solid-red lines corresponds to the odd--transitions. Panel (b) shows a small region of the spectral positions for few inner Rabi doublets around $\lambda=0$ and the color code is the same as panel (a). For interpretation of the references to color in this figure legend, the reader is referred to the web version of this article.}\label{Fig_1}
\end{figure}
Fig.~\ref{Fig_1}(a) shows the  spectral positions for the first inner Rabi doublets ($n=1,2,3,...,19$) as a function of the parity deformation parameter $\lambda$ and the dimensionless energy scale calibrated by the light-matter interaction constant $g$. In such units, the position of the first Rabi doublet is easily identified when $\lambda=0$ at distances of $\pm1$ (horizontal dashed line for guide eye), while that the rest of the higher optical transitions reveal a multiplet structure as is expected in the non-deformed JC model. As the deformation parameter takes positive values, a reduction of the position of the inner Rabi doublets is observed for all of the optical transitions coming from even to odd rungs in the JC ladder (even--transitions). In contrast with the optical transitions coming from odd to even rungs in the JC ladder (odd--transitions) where the position of the inner Rabi doublets increase. Interestingly enough, we find that for the critical value of $\lambda=0.5$ a collective phenomenon emerges in such a way that all even--transitions emit photons at the cavity frequency. This phenomenology also occurs when the deformation parameter decreases until it reaches the critical value of $\lambda\approx-0.5$, but now the collective phenomenon emerges for all odd--transitions which emit photons at the cavity frequency. Fig.~\ref{Fig_1}(b)
shows a small region of the spectral positions for few inner Rabi doublets around $\lambda=0$ (horizontal dashed line for guide eye). Notice that in the case of the non-deformed JC model, the multiplet structure has an alternating parity beginning from odd--transitions (read them from right to left starting with $n=2$, since $n=1$ is not shown in the figure), in contrast to, for example, the case when $\lambda=\pm0.3$ where the multiplet structure has been grouped by parity. In other words, the inner Rabi doublets are separated into two groups of peaks where one of them is coming from even-- or odd--transitions exclusively.
%
\begin{figure}[ht!]
\centering
\includegraphics[scale=1]{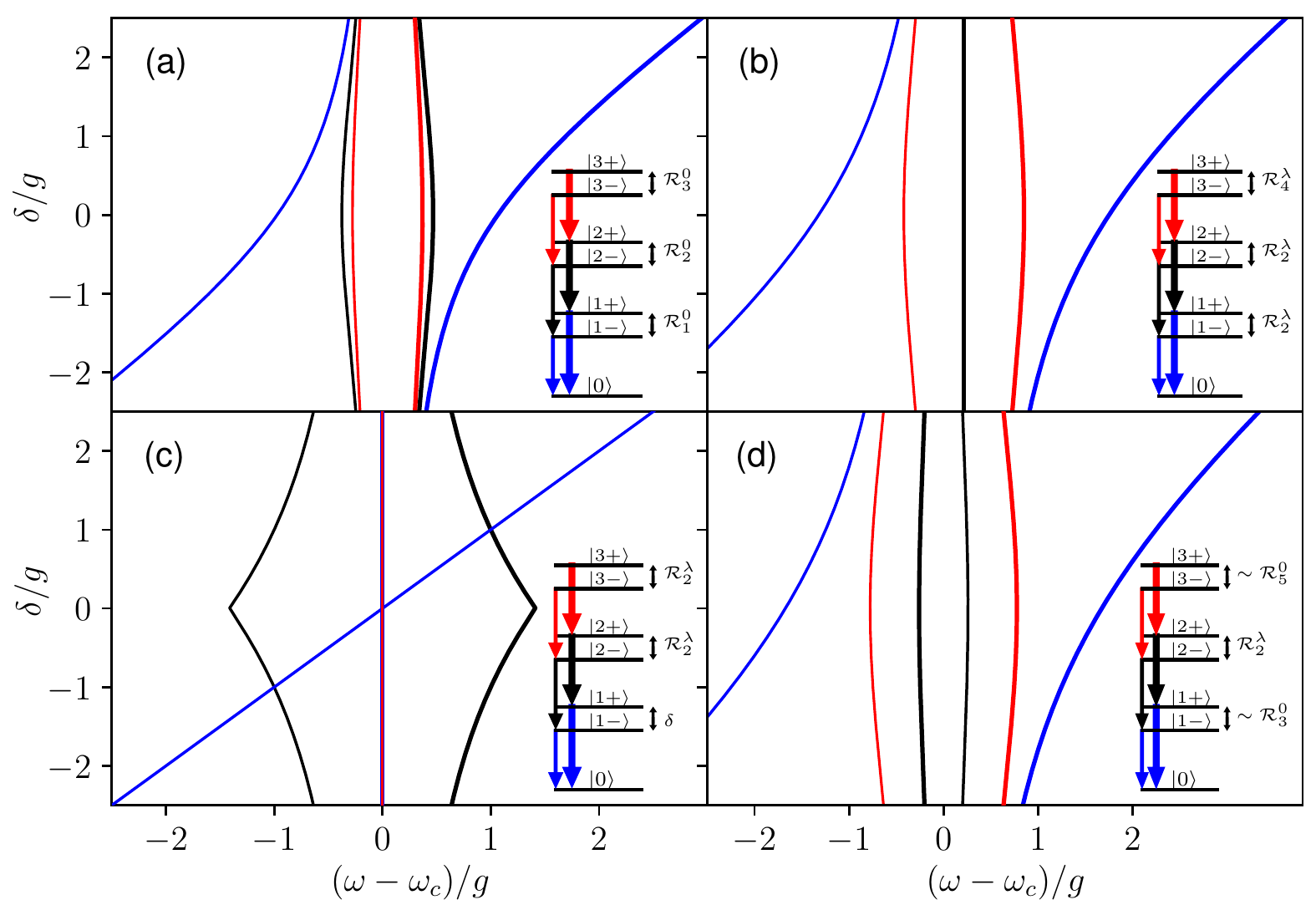}
\caption{Spectral positions of the Rabi doublets $n=1,2,3$ as a function of the detuning $\delta/g$. Panel (a) shows the non-deformed case $\lambda=0$ where the first Rabi doublet shown as solid-blue thick and thin lines. The second Rabi doublet  shown as solid-black thick and thin lines. The third Rabi doublet shown as solid-red thick and thin lines. The inset shows the allowed optical transitions in the JC ladder together with the splitting in energy due to the generalized Rabi frequency $\mathcal{R}^{\lambda}_n$. Similar results are shown in panel (b) but for the deformed case when $\lambda=0.5$, panel (c) when $\lambda=-0.5$ and panel (d) when $\lambda=0.9$. For interpretation of the references to color in this figure legend, the reader is referred to the web version of this article.\label{Fig_2}}
\end{figure}
In Fig.~\ref{Fig_2}(a) we plot the spectral position of first three Rabi doublets as a function of the detuning $\delta$ together with the condition of $\lambda=0$. This particular case corresponds to the well-known non-deformed JC model, where the inner Rabi doublets are separated spectrally by a distance of $2g(\sqrt{n}-\sqrt{n-1})$ at resonance ($\delta=0$) as a signature of the strong coupling regime in the system. Additionally, the inset illustrates the allowed one-photon optical transitions in the JC ladder together with the corresponding generalized Rabi frequency $\mathcal{R}_n^0$ with $n=1,2,3$. At the critical value of $\lambda=0.5$ there is an enlargement of the Rabi splitting for the odd-transitions without losing the signature of the strong coupling regime. In contrast, the collective phenomenon is presented for all even-transitions where the corresponding Rabi doublets merge in a single peak (at the cavity frequency) evidencing signatures of the weak coupling regime as shown in Fig.~\ref{Fig_2}(b). At the critical value of $\lambda\approx-0.5$ there is a level anticrossing as a signature of the strong coupling regime but now originated by the second Rabi doublet in the disguise of the vacuum Rabi doublet. Now the collective phenomenon appears for all odd-transitions, and its corresponding Rabi doublets overlap in a single peak (at the cavity frequency) as in the weak--coupling regime, i.e., as shown in Fig.~\ref{Fig_2}(c). Finally, it is worthwhile noting that after crossing the critical value of $\lambda=0.5$, the inner Rabi doublets associated with even--transitions interchange their spectral positions in contrast to the non-deformed case. It can be seen in Fig.~\ref{Fig_2}(d) for a deformation parameter value of $\lambda=0.9$, where the spectral positions of the Rabi doublet $n=2$ depicted as solid-thick (solid-thin) black line has been shifted to the right (left) hand side. Even though the optical transitions offer valuable information on the spectral position of the emission peaks that could be observed in the PL spectrum, this theoretical approach cannot provide information on the intensity of these emission peaks neither ensure that they can be resolved when the system is interacting with the environment. Therefore, in order to study the emission properties of the deformed-JC model, we must incorporate in our model  irreversible processes such as the leakage of photons from the cavity at the rate $\kappa$, the spontaneous emission of the QE at the rate $\gamma$, as well as the incoherent pumping of the QE at the rate $P$ through the master equation approach as follows:
\begin{equation}\label{exactnumEq}
\frac{d\hat{\rho}}{dt}=-i[\hat{H}_{\lambda},\hat{\rho}]+\frac{\kappa}{2}\mathcal{L}_{\bm{\hat{a}}}(\hat{\rho})+\frac{\gamma}{2}\mathcal{L}_{\hat{\sigma}}(\hat{\rho})+\frac{P}{2}\mathcal{L}_{\hat{\sigma}^\dagger}(\hat{\rho})
\end{equation}
where $\mathcal{L}_{\hat{O}}(\cdot)=(2\hat{O}\cdot\hat{O}^{\dagger}-\hat{O}^{\dagger}\hat{O}\cdot-\cdot\hat{O}^{\dagger}\hat{O}$) defines the Lindblad superoperator for an arbitrary operator $\hat{O}$. Notice that the irreversible processes mentioned above are among the most widely used for describing the interaction of QE-cavity system with an environment. The~\ref{Apendix} shows in detail how the term associated with the irreversible process of leakage of photons from the cavity is obtained within the Lindblad master equation approach. In what follows, it is assumed that the emission originated by the QE is negligible and the PL spectrum becomes mostly from the cavity. Taking into account the Wiener-Khintchine theorem which states a relationship between the correlation function and the power spectrum, it is possible to compute the emitted light by the deformed-JC system as a Fourier Transform of the two-time correlation function of the parity deformed operator field $\bm{\hat{a}}$. More precisely,  $S_{a}(\omega)=\int_{-\infty}^{\infty} \meanv{\bm{\hat{a}}^\dagger(\tau)\bm{\hat{a}}(0)}e^{-i\omega\tau}d\tau$ and for the two-time correlation function should be used the quantum regression formula~\cite{Perea:2004}. Fig.~\ref{fig:PL}(a) shows a comparison of the numerically obtained PL spectrum for the non-deformed JC system (solid-gray line) at resonance with its deformed counterpart for $\lambda=0.5$ (solid-blue line) and $\lambda=-0.5$ (solid-red line). It is well-known that in the non-deformed case, the PL spectrum exhibits the first Rabi doublet together with two inner peaks that are conformed by all one-photon optical transitions ($n>2$). In contrast with this, the deformed case ($\lambda=0.5$) displays a spectral triplet constituted by a central peak which has only contributions from even--transitions and the first Rabi doublet with a mayor splitting. On the other hand, the PL spectrum in the deformed case when $\lambda=-0.5$ displays a spectral triplet where the central peak has only contributions from odd--transitions which are significantly reinforced by the optical transition coming from the first rung, moreover of two satellite peaks formed by the second Rabi doublet. This is complete agreement with the expected results from the optical transitions between the JC ladder (see Fig.~\ref{Fig_1}(a)). Additionally, the Fig.~\ref{fig:PL}(b) compares the PL spectrum of the system between the deformed case with $\lambda=0.9$ (solid-green line) and the non-deformed case (solid-gray line) for a detuning of $\delta/g=0.7$. 
Notice that, even though these PL spectra share similar features as is the presence of four peaks, where the external ones correspond to the first Rabi doublet, the internal peaks in the deformed case comes from even-transitions exclusively, in contrast to the non-deformed case where the internal peaks come from odd- and even-transitions in the JC ladder. \\
\begin{figure}[ht!]
\centering
\includegraphics{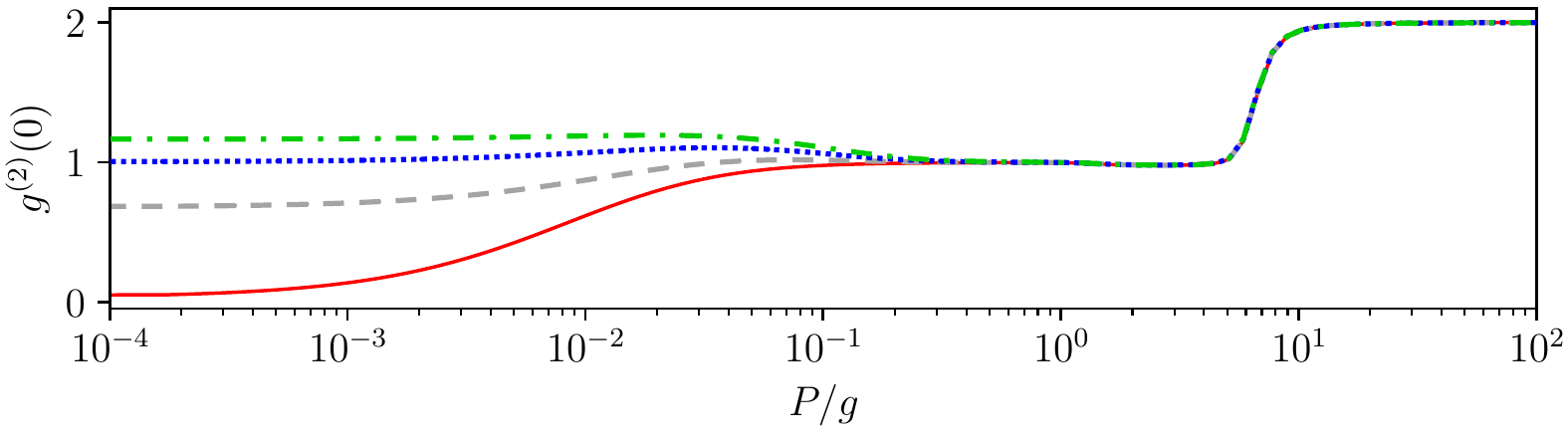}
\caption{Steady-state second-order correlation as a function of pumping $P/g$ for different values of the deformation parameter $\lambda$. Dashed-gray line shows the non-deformed case, solid-red line shows the case when $\lambda=-0.5$ and the dotted-blue line shows the case when $\lambda=0.5$. Finally, dot-dashed green line shows the case when $\lambda=0.9$. The parameters are fixed to
$\kappa/g=0.083$, $\gamma/g=0.017$ and
$\delta/g=0$.
}\label{fig:g2}
\end{figure}
%
It is well-known that there is special interest in research on semiconductor single-photon sources (SPs) from experimental side and important efforts have been made on the measurements of optical correlation functions. The second-order correlation function is considered as a valuable criterion for the characterization of the emitted light by QE-cavity systems, since they could be quantum devices operating as SPs being the fundamental blocks in both computing and quantum information processing. In order to investigate the effect of the parity deformation parameter on the emitted light by the QE-cavity system. We compute the steady-state second-order correlation function at zero time delay through the expression  $g^2(0)=\meanv{\hat{\mathbf{a}}^\dagger(0)\hat{\mathbf{a}}^\dagger(0)\hat{\mathbf{a}(0)}\hat{\mathbf{a}}(0)}/\meanv{\hat{\mathbf{a}}^\dagger(0)\hat{\mathbf{a}}(0)}$.
In particular, the Fig.~\ref{fig:g2} shows the steady-state second-order correlation function as a function of the pumping $P$. We observe that the parity deformation parameter plays an important role at low pumping rate, whereas at high pumping rate, there are no differences between non-deformed respect to the deformed case. At this range of pumping, the emitted light corresponds to the well-known description of classical light ($g^2(0)=2$). It is well-known that in the non-deformed case, the second-order correlation function is smaller than one but larger than zero, and it is attributed to non-perfect anti-bunching in the system (dashed-gray line). Interestingly, we observe that when the parity deformation parameter is set to $\lambda=0.5$ the QE-cavity system operates as a source of coherent light since $g^2(0)\approx1$ (dotted-blue line) for all range of the parameter $P$. When the parity deformation parameter is set to $\lambda=-0.5$ the light emitted from the QE-cavity system is perfectly anti-bunched, since $g^2(0)\approx0$ (solid-red line). Surprisingly, for this particular value of the parity deformation parameter $g^2(0)<1/2$, which implies that the system represents a good single-photon light source~\cite{Grunwald:2019}.
Finally, at $\lambda=0.9$ it is found that the statistical properties of the emitted light by the system corresponds to a classical field, as shown in the dot-dashed green line. Our results reveal that the parity deformation parameter changes the emission properties of light in the QE-cavity system without requiring significant external excitation.
\section{Conclusions}\label{sec:conclusions}
\noindent
We have investigated a pumped-dissipative QD-cavity system within the Lindblad master equation approach together with the RDHA framework. In particular, we have studied the dependence of the parity deformation parameter on one-photon optical transitions of the system, and it was also demonstrated that this parameter induces new phenomenology in the emission spectrum of the system. Our findings have shown that depending on the critical value of the deformation parameter, a spectral triplet emerges in the PL spectrum that exhibits properties of the weak- and strong-coupling regimes. More precisely, an emission spectrum that consists of a single peak at the resonance (cavity frequency) as in the weak coupling and a double peak that emerges in the emission spectrum, arising from the QE-cavity level anticrossing as in the strong coupling. Finally, we have shown that for those critical values of the deformation parameter together with a low pumping regime, the system can operate as a good single-photon light source or as a coherent source of light.
\section*{Acknowledgments}
\noindent
S.E.-A. gratefully acknowledges funding by COLCIENCIAS from ``Beca de Doctorados Nacionales'' -- Convocatoria~727--. E.A.G acknowledges financial support from Vicerrector\'ia de Investigaciones at Universidad del Quind\'io.
\appendix
\section{Derivation of the superoperator $\mathcal{L}_{\hat{\mathbf{a}}}$}\label{Apendix}
A commonly used approach for obtaining the different decay rates appearing in a master equation is to assume a microscopic model for the interaction of the system under study and the environment. Within this framework, each one of the sub-systems (either the cavity mode or the QE) interacts individually with an uncorrelated local environment and gives origin to the distinct Lindblad terms that appear in the master equation given by Eq.~(\ref{exactnumEq}). In consequence, each one of the irreversible processes is incorporated into the master equation through the sum of their respective Lindblad superoperators. In what follows, we show how the superoperator term associated with leakage of photons from the cavity at a rate $\kappa$ is obtained. Let us consider a thermal bosonic environment defined through the Hamiltonian given by ($\hbar=1$)
\begin{equation}
\hat{H}_{E}=\sum_i\Omega_i\hat{b}_i^\dag\hat{b}_i
\end{equation}
where $\hat{b}_i^\dag$ and $\hat{b}_i$ define the creation and annihilation operators for $i$-th mode of the thermal environment with frequency $\Omega_i$, respectively. Additionally, we assume that the environment interacts with the cavity mode via a dipolar interaction as follows 
\begin{equation}
\hat{H}_{EC}=\sum_{i}{g_{i}(\hat{b}_{i}^\dag \hat{\mathbf{a}} + \hat{b}_{i} \hat{\mathbf{a}}^\dag)},
\end{equation} 
where $g_i$ defines the system-environment coupling constant and assumed to be real. Additionally, the Hamiltonian describing the system-environment is given by
\begin{equation}
\hat{H}_{SE}=\hat{H}_\lambda+\hat{H}_{E}+\hat{H}_{EC}\label{HSe}    
\end{equation}
and where the Hamiltonian given by Eq.~(\ref{JCmodel}) has been rewritten in a more convenient form as follows:
\begin{equation}
\hat{H}_\lambda=\hat{H}_S+\hat{H}_{int},\label{H}
\end{equation}
where 
\begin{equation}
\hat{H}_S=\frac{\omega_x}{2}\hat{\sigma}_z+\frac{\omega_c}{2}(1+2\lambda+2\hat{N})
\end{equation}
and
\begin{equation}
\hat{H}_{int}=g(\mathbf{\hat{a}}^{\dagger}\hat{\sigma}+\mathbf{\hat{a}}\hat{\sigma}^{\dagger}).
\end{equation}
It is straightforward to move into the interaction picture using a unitary transformation with $\hat{H}_S+\hat{H}_E$ as follows:
\begin{eqnarray}
\hat{H}(t)&=&e^{i(\hat{H}_S+\hat{H}_{E})t}\hat{H}_{\lambda}e^{-i(\hat{H}_S+\hat{H}_{E})t},\\\hat{H}(t)&=&e^{i(\hat{H}_S+\hat{H}_{E})t}\hat{H}_{\lambda}e^{-i(\hat{H}_S+\hat{H}_{E})t},\\
\hat{H}_I(t)&=&e^{i(\hat{H}_S+\hat{H}_{E})t}(\hat{H}_{E}+\hat{H}_{EC})e^{-i(\hat{H}_S+\hat{H}_{E})t}\label{hint},\\
\hat{\rho}_I(t)&=&e^{i(\hat{H}_S+\hat{H}_{E})t}\hat{\rho}_{SE}e^{-i(\hat{H}_S+\hat{H}_{E})t}.
\end{eqnarray}
where $\hat{\rho}_{SE}$ denotes the joint system-environment density operator. It is worth mentioning that the unitary transformation given by Eq.~(\ref{hint}) can be handled as in a canonical derivation of a Lindblad master equation, since the involved commutators $[\hat{N},\hat{\textbf{a}}]$,  $[\hat{N},\hat{\textbf{a}}^\dagger]$, $[\hat{H}_S,\hat{\textbf{a}}]$, $[\hat{H}_S,\hat{\textbf{a}}^\dagger]$ follow the Baker-Hausdorff lemma~\cite{Sakurai:book} as its non-deformed counterpart and they do not introduce additional terms. Then, the Liouville-von Neumann equation of $\hat{\rho}_I(t)$ is given by
\begin{equation}
\frac{d\hat{\rho}_I(t)}{dt}=-i[\hat{H}(t),\hat{\rho}_I(t)]-i[\hat{H}_I(t),\hat{\rho}_I(t)].
\end{equation}
In what follows, we consider the well-established Born-Markov approach and perform the partial trace over the environment degrees of freedom for obtaining \begin{equation}\label{aux2}
\frac{d\hat{\rho}_I}{dt}=-i[\hat{H}(t),\hat{\rho}_I(t)]-i\meanv{[\hat{H}_{I}(0),\hat{\rho}_I(0)]}-\int_0^\infty dt'\meanv{[\hat{H}_{EC}(t),[\hat{H}_{EC}(t'),\hat{\rho}_I(t)]]}_{E}.
\end{equation}
From the separability approximation at $t=0$, the second term on the right-hand side of Eq.~(\ref{aux2}) vanishes and the third term can be expanded to obtain:
\begin{eqnarray}\label{din_mat_4}
\frac{d\hat{\rho}_I}{dt} &=&-i[\hat{H}(t),\hat{\rho}_I(t)]\notag\\
&&-\int_0^\infty dt'\sum_{j=1}^{2}\Big \{[(\hat{\mathbf{a}}^\dag \hat{\mathbf{a}}\hat{\rho}_I(t)-\hat{\mathbf{a}}\hat{\rho}_I(t) \hat{\mathbf{a}}^\dag)e^{-i\left(\omega_c-\omega_i\right)(t-t')}]\meanv{\sum_i g_i^2 \hat{b}_i^\dag \hat{b}_i e^{i\Omega_i(t-t')}}_{E}\notag \\
&&+ [(\hat{\rho}_I(t) \hat{\mathbf{a}} \hat{\mathbf{a}}^\dag - \hat{\mathbf{a}}^\dag \hat{\rho}_I(t) \hat{\mathbf{a}})e^{-i\left(\omega_c-\omega_i\right)(t-t')}]\meanv{\sum_i g_i^2 \hat{b}_i\hat{b}_i^\dag e^{i\Omega_i(t-t')}}_{E}\notag \\
&&+[( \hat{\mathbf{a}}\hat{\mathbf{a}}^\dag \hat{\rho}_I(t)- \hat{\mathbf{a}}^\dag \hat{\rho}_I(t) \hat{\mathbf{a}})e^{i\left(\omega_c-\omega_i\right)(t-t')}]\meanv{\sum_i g_i^2 \hat{b}_i\hat{b}_i^\dag e^{-i\Omega_i(t-t')}}_{E}\notag \\
&&+ [(\hat{\rho}_I(t) \hat{\mathbf{a}}^\dag \hat{\mathbf{a}} - \hat{\mathbf{a}} \hat{\rho}_I(t) \hat{\mathbf{a}}^\dag)e^{i\left(\omega_c-\omega_i\right)(t-t')}]\meanv{\sum_i g_i^2 \hat{b}_i^\dag \hat{b}_i e^{-i\Omega_i(t-t')}}_{E} \Big \},
\end{eqnarray}
where the partial trace over the degrees of freedom of the environment is denoted by the symbol $\meanv{\cdot}_{E}=Tr_{E}[\cdot\hat{\rho}_{I}]$. It is assumed to be thermal. Then, they can be cast in:  
\begin{eqnarray}
\meanv{\sum_i g_i^2 \hat{b}_i^\dag \hat{b}_i e^{i\Omega_i(t-t')}}_{E}&=&\sum_ig_i^2e^{i\Omega_i(t-t')}\bar{n},\\ 
\meanv{\sum_i g_i^2 \hat{b}_i\hat{b}_i^\dag e^{i\Omega_i(t-t')}}_{E}&=&\sum_ig_i^2e^{i\Omega_i(t-t')}(1+\bar{n}),\\
\meanv{\sum_i g_i^2 \hat{b}_i\hat{b}_i^\dag e^{-i\Omega_i(t-t')}}_{E}&=&\sum_ig_i^2e^{-i\Omega_i(t-t')}(1+\bar{n}),\\
\meanv{\sum_i g_i^2 \hat{b}_i^\dag \hat{b}_i e^{-i\Omega_i(t-t')}}_{E}&=&\sum_ig_i^2e^{-i\Omega_i(t-t')}\bar{n},
\end{eqnarray}
where $\bar{n}(\omega)=[e^{\omega/T}-1]^{-1}$ is the Bose-Einstein distribution at a temperature $T$ (the Boltzmann's constant $\kappa_B$=1). Finally, it is introduced the spectral density of the environment $D(\omega)$ and taking into account the continuum as $\sum_i\to\int d\omega D(\omega)$, we go back to to the Schr\"odinger representation in the Eq.~(\ref{din_mat_4}). Now the master equation reads explicitly as follows:
\begin{equation}
\frac{d\hat{\rho}}{dt} =-i[\hat{H}_\lambda,\hat{\rho}]+ \frac{\pi}{2} g(\omega_c)^2D(\omega_c)(1+\bar{n})\mathcal{L}_{\hat{\mathbf{a}}}(\hat{\rho})+ \frac{\pi}{2} g(\omega_c)^2D(\omega_c)\bar{n}\mathcal{L}_{\hat{\mathbf{a}}^\dagger}(\hat{\rho}).
\end{equation}
Thus, the Eq.~(\ref{exactnumEq}) is recovered by taking $\bar{n}=0$ and defining the photonic decay rate as $\kappa=\pi g(\omega_c)^2D(\omega_c)(1+\bar{n})$. Here $g(\omega_c)$ defines the system-environment interaction at the frequency $\omega_c$. 

\end{document}